\documentclass{aa}
\usepackage{graphicx}
\begin{document}
   \title{The complex iron line of NGC~5506}


   \author{G. Matt
          \inst{1}
          \and
          M. Guainazzi\inst{2}
	  \and
	  G.C. Perola\inst{1}
        \and    F. Fiore \inst{3}
        \and    F. Nicastro \inst{4}
        \and    M. Cappi \inst{5}
	\and L. Piro \inst{6}
          }

   \offprints{G. Matt}   

\institute{Dipartimento di Fisica, Universit\`a degli Studi Roma Tre,
              Via della Vasca Navale 84, I--00146 Roma, Italy
   \and XMM-Newton SOC, VILSPA--ESA, Apartado 50727, E--28080 Madrid, Spain
   \and Osservatorio Astronomico di Roma, Via dell'Osservatorio,
        I--00044 Monteporzio Catone, Italy
\and  Harvard-Smithsonian Center of Astrophysics, 60 Garden Street,
Cambridge MA 02138 USA
\and Istituto Tecnologie e Studio Radiazioni Extraterrestri, CNR, Via
              Gobetti 101, I--40129 Bologna, Italy
\and Istituto di Astrofisica Spaziale, C.N.R., Via Fosso del Cavaliere,
                I--00133 Roma, Italy
   }

   \date{Received; accepted  }

   \abstract{The bright Narrow Emission Line Galaxy, NGC~5506, has been
observed simultaneously by XMM--$Newton$ and BeppoSAX. The iron line 
is complex, with at least two components: one narrow and corresponding
to neutral iron, the second one broad and corresponding to ionized iron.
The latter line is equally well fitted by a truly broad line or by a blend
of He-- and H--like narrow lines. The bulk of the Compton reflection continuum
originates in neutral matter, and is therefore associated with the narrow
line: they are most likely emitted in distant matter. The origin
of the ionized line(s) is less certain, but the solution in terms of a blend
of narrow lines from photoionized matter seems to be preferable to emission 
from an ionized, relativistic accretion disc.
   \keywords{Galaxies: individual: NGC~5506 --
                Galaxies: Seyfert --
                X-rays: galaxies }
               }

   \maketitle

\section{Introduction}

Recent $Chandra$
(e.g. Kaspi et al. 2001; Yaqoob et al. 2001) and XMM--$Newton$ (e.g
Reeves et al. 2001; Pounds et al. 2001; Gondoin et al. 2001)
results clearly indicate that a narrow iron line
component is often, if not always, present in Seyfert 1 galaxies, alone or
together with the relativistic disc component (Fabian et al. 2000 
and references therein). 
An analysis of the composite $ASCA$ spectrum of Seyfert 1s
(Lubinski \& Zdziarski 2001) also suggests the presence of a narrow
component, confirming previous findings on individual sources. 
Whether this component
arises in Compton--thin matter like the Broad Line Region or the Narrow Line 
Region, or in 
Compton--thick material 
like the `torus' (Antonucci 1993) or the outflowing matter (Elvis 2000)
envisaged in Unification Models 
and clearly present in several Seyfert 2s (Maiolino et
al. 1998; Risaliti et al. 1999; Matt 2001 and references therein),  is however
still a matter of debate. 
The high energy resolution $Chandra$ observation
of NGC~5548 (Yaqoob et al. 2001)
just resolved the line width, which comes out to be consistent
with those of the optical broad lines. The upper limit on the line width 
in NGC~3783 (Kaspi et al. 2001) is instead 
consistent with the iron line originating outside the BLR. 

In this paper we present the
results from a XMM--$Newton$ observation of NGC~5506 {\sl simultaneous}
with a BeppoSAX observation, aimed to study in unprecedented detail
the reprocessed components by combining the sensitivity
of XMM--$Newton$ at the iron line energy with the still unique capability
of BeppoSAX in hard X--rays.  

The Narrow Line Emission Galaxy NGC~5506 is one of the brightest AGN
in hard X--rays, and for this reason has been
extensively studied in the past. The nucleus is obscured by cold matter
with a column density of about 3$\times$10$^{22}$ cm$^{-2}$.
A possibly variable soft component
(Bond et al. 1993) was discovered by $GINGA$, which also detected the
iron K$\alpha$ line and the reflection component. Both $ASCA$ spectroscopic
(Wang et al. 1999) and $RXTE$ variability (Lamer et al. 2000) 
observations suggest that the line is complex.
BeppoSAX (Perola et al., 2001)
found an iron line centroid energy bluer than 6.4 keV, i.e. 
E$_{\rm k}$=6.52$\pm$0.09 (equivalent width of 150$\pm$40 eV) and measured
the reflection component at $R$=1.2$\pm$0.4. Only a lower limit
of 300 keV could be put on the high--energy exponential cut--off. 

\section{Observations and data analysis}

XMM--$Newton$ (Jansen et al. 2001) observed NGC~5506 between February 2
(17:34 UT) and February 3 2001 (03:35 UT).  Imaging CCD cameras (EPIC-MOS,
0.2-10~keV, Turner et al. 2001; EPIC-p-n, 0.2--15~keV,
Str\"uder et al. 2001) were operated in
Large Window mode, with the Medium filter.
High resolution spectroscopy cameras (RGS; 0.2--1.5~keV;
der Herder et al. 2001)
were simultaneously
operating, but -- due to the large absorbing column --
they detected the source only at the $\simeq$4$\sigma$
level. No evidence for narrow absorption or emission
structure is present in the RGS spectra, and we will not deal
with these data any longer in this paper.
Data were reduced with {\sc SAS v.5.0} (Jansen et al. 2001),
using the  calibration files publicly available on April 2001.
After data screening the total exposure time was 17.6~ks and
13.8~ks for the MOS and the p-n, respectively.
X-ray events corresponding to pattern 0-12 for the MOS 
and to pattern 0 for the p-n
were used. Non X-ray background remained low throughout the
observation. Spectra and light curves were extracted in regions of 40'' radii.
The p-n count rate in the 0.1--15~keV band is $3.75 \pm 0.16$, corresponding
to a pile-up fraction of $\simeq$0.5\% only (in the MOS the pile-up fraction is
about 0.7\%). In particular, the p-n count rate in the 6--7 keV range
is 3 times than in ASCA/SIS, making it the best CCD instrument 
ever for iron line studies. 

BeppoSAX observed the source from February 1 (5:50 UT) to February 3
(13:20 UT), 2001, for a total net exposure time of 78 ks for the MECS
and PDS (2 units) instruments, and 29 ks for the LECS. Data reduction
and analysis were standard (e.g. Guainazzi et al. 1999). LECS and MECS
spectra and light curves were extracted in regions of 4' radii. 

The source
varied in flux by about 40\% during the BeppoSAX observation, and by about
20\% during the XMM--$Newton$ observation. No spectral variability
has been detected.

In this paper we used only the XMM/EPIC-p-n and BeppoSAX/PDS instruments
after having checked
that the spectra from XMM/EPIC-MOS and BeppoSAX/MECS instruments
were consistent with that of XMM/EPIC-p-n. (A temporal and
spectral analysis of the complete XMM--$Newton$ and BeppoSAX data sets
is beyond the scope of this paper and is deferred to a future work). 
In order to maximize the statistics, and because of the lack of spectral
variability,
we integrated the PDS spectrum over the entire observation despite only a part
of it was covered by the XMM--$Newton$ observation.
In fitting the p-n and 
PDS spectra we introduced a multiplicative factor of 1.215 for the PDS
to take into account cross--calibrations 
and the different exposure times. This factor was obtained by
normalizing the p-n flux to the integrated MECS spectrum 
and applying a relative normalization factor of 0.84 between the
BeppoSAX PDS and MECS (Fiore et al. 1999). 

Spectral analysis has been performed with the {\sc xspec v.11} software
package.
All errors refer to 90\% confidence level for 1 interesting parameter
($\Delta\chi^2$=2.7). 

\section{Results} 

\subsection{The Baseline model}

   \begin{figure}
   \centering
   \includegraphics[angle=-90,width=8.5cm]{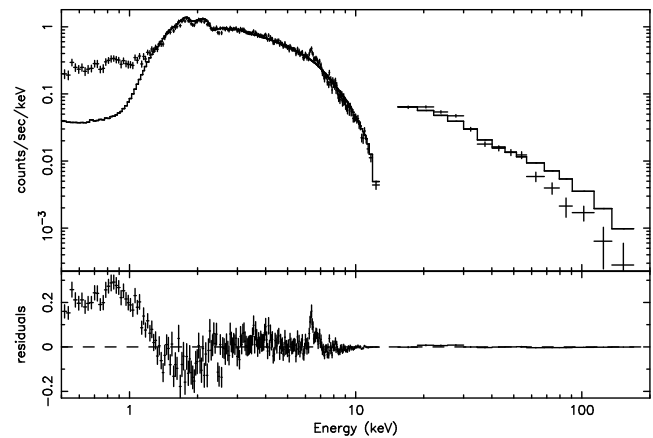}
      \caption{The EPIC--p-n and BeppoSAX--PDS spectra fitted with a simple
absorbed power law. }
         \label{badfit}
   \end{figure}

   \begin{figure}
   \centering
   \includegraphics[angle=-90,width=8.5cm]{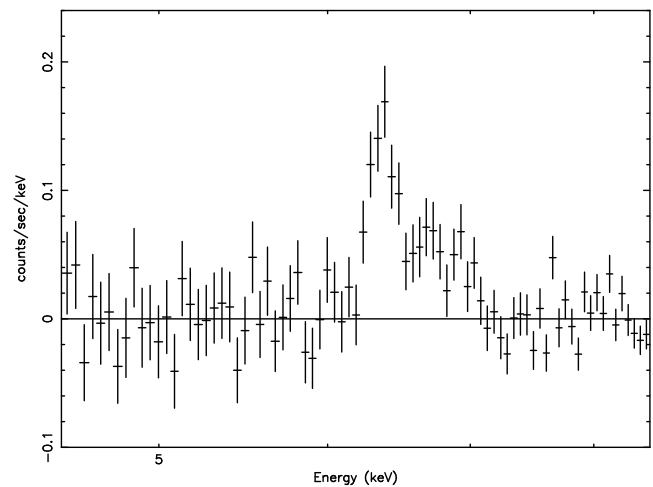}
      \caption{ Residuals in the iron line region. At least two iron line
components are clearly present. }
         \label{line}
   \end{figure}

In Fig.~1 the p-n/PDS spectrum is shown, together with the residuals 
after fitting the data with a simple absorbed power-law
model. A soft excess and a strong iron line are 
clearly apparent in the residuals.  The high energy
part of the spectrum is also badly fitted.
A blow--up of the residuals around the iron line
shows clearly that the line is complex, being composed by a narrow
component around 6.4 keV and a broader component bluewards.

Because in this paper
we are mainly interested in the reprocessed components, for the sake
of simplicity we excluded from
the analysis the energies below 2.5 keV, where the soft excess is present.

We first added to the model two gaussian lines. One of the lines turns out
to be unresolved and close to 6.4 keV, 
the second one broad and corresponding to ionized iron.
The quality of the fit improves, but it is still unacceptable 
($\chi^2_{\rm r}$/d.o.f.=243.9/173). 
Because most of the $\chi^2$ comes from the highest energy part of
the spectrum, we added also a reflection component
({\sc pexrav} model: Magdziarz \& Zdziarski 1995). 
The fit is now perfectly acceptable: $\chi^2_{\rm r}$/d.o.f.=176.4/172.
(We note here that a fit with only one gaussian line, which basically fits the
narrow component, gives 
$\chi^2_{\rm r}$/d.o.f.=202.9/175. The ionized line is therefore significant
at more than 99.99\% confidence level, according to the F--test.)
The best fit parameters are summarized in Table~1.

The fit without the PDS gives similar values
for the iron lines and the power law
index, but only an upper limit to $R$ of 1.6.

To be sure that the above findings do not depend on the limited energy
band adopted, we repeated the above analysis extending the energy band down
to 0.5 keV. The soft excess is well fitted by a partial covering model
with covering fraction of about 98\%, plus a further, complete absorber
with column density of about 2$\times10^{21}$ cm$^{2}$. Again, the inclusion
of a second gaussian line is highly significant
($\chi^2_{\rm r}$/d.o.f.=318.9/250 for one line, and 289.5/247 for two
lines; the second line is significant at more than 99.99\% confidence level). 
The best--fit parameters are all consistent within the errors with those 
reported in Table~1. 

\begin{table}
\caption{Baseline Model Spectrum fit ($i$ is fixed to 30$^{\circ}$ in the CR
model). Line energies are in the source rest frame ($z$=0.0061).}

\vspace{0.05in}
\begin{tabular}{|c|c|}
\hline
~ & ~ \cr
F(2-10~keV) (erg~cm$^{-2}$~s$^{-1}$) & 7.8$\times10^{-11}$ \cr
$\Gamma$ & 1.98$^{+0.05}_{-0.02}$ \cr
N$_H$~(10$^{22}$~cm$^{-2}$) & 3.44$^{+0.13}_{-0.12}$ \cr
E$_f$~(keV) & 330$^{+120}_{-80}$ \cr
$R$ & 1.09$^{+0.08}_{-0.08}$ \cr
E$_{k,1}$~(keV) & 6.41$^{+0.03}_{-0.03}$ \cr
$\sigma_{k,1}$~(keV) & $<0.06$ \cr
I$_{k,1}$~(ph~cm$^{-2}$~s${-1}$) & 5.0$^{+2.2}_{-0.7}\times$10$^{-5}$ \cr
EW$_{k,1}$~(eV) & 70$^{+30}_{-10}$ \cr
E$_{k,2}$~(keV) & 6.75$^{+0.10}_{-0.15}$ \cr
$\sigma_{k,2}$~(keV) & 0.25$^{+0.08}_{-0.06}$ \cr
I$_{k,2}$~(ph~cm$^{-2}$~s${-1}$) & 7.0$^{+1.8}_{-1.9}\times$10$^{-5}$ \cr
EW$_{k,2}$~(eV) & 110$\pm$30 \cr
$\chi^2$/d.o.f. & 176.4/172 \cr
~ & ~ \cr
\hline
\end{tabular}
\end{table}

\subsection{The origin of the iron lines}

The 6.4 keV iron line is clearly too narrow to come from the innermost
part of the accretion disc. It may arise either in Compton--thin material,
like the Broad Line Region or Narrow Line Region, 
or in Compton--thick matter, like the outermost
part of the accretion disc or the `torus'. To distinguish between
the two possibilities it is necessary to understand whether the Compton
Reflection (CR) component is at least partly 
associated with the 6.4 keV line (so indicating emission from Compton--thick
matter) or, instead, is completely associated with 
the ionized line (in which case the matter responsible
for the narrow line must be Compton--thin). 

We first tested whether the CR may be completely associated with the ionized
line, supposing that both are emitted in a  relativistic disc. We 
accounted for relativistic effects in both the line and CR by adopting
the models {\sc diskline} (Fabian et al. 1989) and {\sc refsch}, respectively.
All disc parameters were forced to be the same in the  
{\sc diskline} and {\sc refsch}
models. We fixed the inner radius to 6$r_{\rm g}$ (the innermost stable
orbit in Schwarzschild metric) and left the outer radius and the inclination
angle as free parameters. (For simplicity, 
here and in all following fits, the energy of the unresolved line has been 
fixed to 6.4 keV, and its width to zero.) The fit is good 
($\chi^2$/d.o.f.=174.6/172) but the results not physically
self--consistent: in fact,
while the line rest frame energy confirms that the iron must be ionized
(E$_{\rm k}$=6.78$\pm$0.08 keV), the ionization parameter of the reflecting
matter is very low ($\xi<$0.17 erg cm s$^{-1}$) and definitely inconsistent 
with the ionization inferred from the iron line. 

After fixing, for simplicity, the energy
of the iron line to 6.7 keV, corresponding to He--like iron,
we forced the disc
to be really ionized, i.e. with the ionization parameter not less than
500 erg cm s$^{-1}$
(to be consistent with the centroid energy of the ionized iron line).
The fit is totally unacceptable ($\chi^2$/d.o.f.=339.9/173; see Fig.~3), 
due to a bad fitting of the continuum at low energies and 
to a deep ($\tau\sim0.2$) edge at 7.1 keV, which in the previous fits were
accounted for by the neutral reflection continuum.
We checked whether this edge may be related to 
the cold absorber by allowing the 
iron abundance to vary with respect to the other elements.
A much better fit ($\chi^2$/d.o.f.=216.0/172)
is found with $A_{\rm Fe}\sim10$, but
still significantly worse than for the baseline model discussed above.
Moreover, the iron overabundance seems unrealistically large. 
Therefore, we conclude that {\it most if not all
of the CR comes from neutral matter and must therefore be
associated with the cold, narrow iron line}. The ratio between the neutral
iron line equivalenth width and the amount of Compton reflection component
is somewhat lower than expected (an EW of about 130 eV is predicted
if $R$=1, $\theta$=30$^{\circ}$ and $\Gamma$=2, 
George \& Fabian 1991 and Matt et al. 1991), suggesting a possible iron
underabundance.

The nature of the ionized line is less clear.
The disc solution is possible but rather unlikely.
Including in the model both a cold and an ionized CR,
a good fit ($\chi^2$/d.o.f.=174.7/172) is found, but
with the cold reflection largely dominating
($R_{\rm cold}$=1.35$^{+0.50}_{-0.25}$, $R_{\rm ion}<0.28$), 
implying a rather large ratio
 between the ionized iron line EW ($\sim$100 eV) 
and the corresponding CR. 
Such a ratio is indeed possible when iron is mainly
in the He--like stage (Matt et al.
1993, 1996; Ballantyne et al. 2001; Nayakshin \& Kallman 2001);
the small value of the CR could be due to either an almost edge--on disc
or a small emitting region. The 
inclination angle of the disc is not well constrained 
in this fit, any value above 25$^{\circ}$ being possible.
However, because the ionized line is not very broad, the best fit value for the
outer radius is pretty large (lower limit of about 300$R_{\rm g}$),
while any realistic disc model suggests strong ionization only in the 
innermost regions. 

Alternatively, the line may be a blend
of two or more narrow lines. A fit with a blend of He--like and H--like
lines is as good as that with the broad line ($\chi^2$/d.o.f.=172.9/175). 
The EW of the two lines are 40$\pm$16 eV and 32$\pm$15 eV, respectively. 
(The EW of the neutral line is now 90$\pm$15 eV, 
other parameters similar to those in Table~1). 
These ionized lines can be produced by fluorescence and resonant scattering
in photoionized matter (Matt et al. 1996). Equivalent widths similar to
the observed ones may be obtained if the Thomson 
optical depth of the ionized matter is a few hundredths (Bianchi 
et al., in preparation). Interestingly, the normalization 
of the soft excess is about 2\% of the primary component.  It is then possible
that the soft X--ray emission and the ionized lines come from one
and the same matter.

   \begin{figure}
   \centering
   \includegraphics[angle=-90,width=8.5cm]{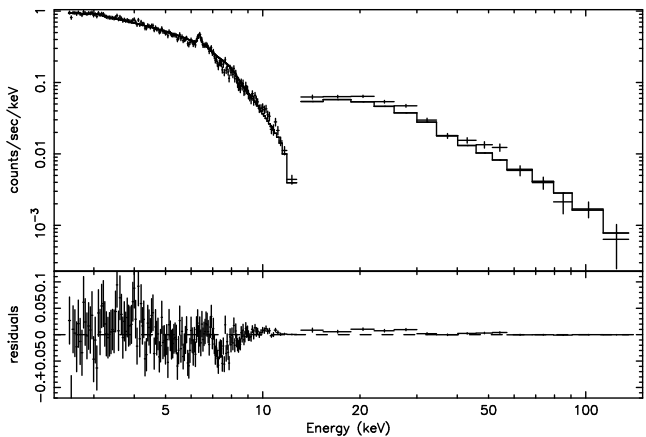}
      \caption{Spectra and best-fit model with only the ionized reflection 
component. See text for details. 
              }
         \label{line}
   \end{figure}

\section{Discussion and Conclusions}

The iron line in NGC~5506 is clearly complex. The cold and narrow
component is associated with the bulk of the Compton
Reflection emission and therefore originates in Compton--thick matter,
like the torus. In NGC~5506 absorption is instead Compton--thin. This implies
either that the torus is dishomogeneous, or that the reflector and
the absorber are different altogether (e.g. Matt 2000). 
This is one of the few cases in which distant, Compton--thick
matter is observed {\it in reflection only}. (The best case so far was
NGC~4051, Guainazzi et al. 1998). If this component will be found to be common
in type 1 Seyfert, this will support one of the least tested predictions
of unification models, i.e. that any Seyfert 1 is surrounded by circumnuclear,
optically thick matter. 

The origin of the ionized component is less
clear, but a solution in terms of a blend of narrow He-- and H--like
iron lines seems preferable to that of a relativistic, ionized disc.
If this is indeed the case, we are left with 
no evidence whatsoever for disc emission, a rather puzzling situation. 
Of course, a trivial possibility is that the disc is nearly edge--on,
reducing the reprocessed components to invisibility. This solution
can certainly work for a single source, but would be untenable if
the same situation would occur in many other sources. 

The iron line complex observed in NGC~5506 is very similar to those
observed by XMM--$Newton$ 
in Mrk~205 (Reeves et al. 2001) and Mrk~509 (Pounds et al. 2001).
One cannot help wondering if this is not the rule for Seyfert
galaxies. However, in the XMM--$Newton$ observation of 
Fairall~9 (Gondoin et al. 2001) only the narrow and cold component is
detected. Moreover, NGC~5506 and Mrk~509
are the only two sources in the BeppoSAX sample of Perola et al. (2001) 
in which the iron line centroid energy is higher than 6.4 keV. It is
therefore possible that these sources represent the exception rather
than the rule.

\begin{acknowledgements}
We thank the BeppoSAX Scientific Data Center 
and the User Support Group of the XMM--$Newton$ SOC for their help.
GM, GCP and FF acknowledge financial support from ASI and from MURST (under
grant {\sc cofin-00-02-36}).
\end{acknowledgements}

\end{document}